\documentclass[11pt]{cernrep}
\def\BA{\begin{eqnarray}}
\def\BE{\begin{equation}}
\def\EA{\end{eqnarray}}
\def\EE{\end{equation}}
\def\eps{\varepsilon}
\def\Dtau{\Delta\tau}
\def\absP{\langle\vert P\vert\rangle}
\def\gtsim{\lower-0.45ex\hbox{$>$}\kern-0.77em\lower0.55ex\hbox{$\sim$}}
\def\ltsim{\lower-0.45ex\hbox{$<$}\kern-0.77em\lower0.55ex\hbox{$\sim$}}

\begin{document}
\title{GEOMETRICAL SCALING DUE TO CRITICAL BEHAVIOR NEAR THE LIGHT CONE}
\author{Hans J. Pirner}
\institute{${}^a$Institut f\"ur Theoretische Physik der
Universit\"at Heidelberg, Germany\\
${}^b$ Max-Planck-Institut f\"ur Kernphysik Heidelberg, Germany }

\maketitle

\def\BA{\begin{eqnarray}}
\def\BE{\begin{equation}}
           \def\EA{\end{eqnarray}}
\def\EE{\end{equation}}
\def\eps{\varepsilon}
\def\Dtau{\Delta\tau}
\def\absP{\langle\vert P\vert\rangle}
\def\gtsim{\lower-0.45ex\hbox{$>$}\kern-0.77em\lower0.55ex\hbox{$\sim$}}
\def\ltsim{\lower-0.45ex\hbox{$<$}\kern-0.77em\lower0.55ex\hbox{$\sim$}}

\begin{abstract}
At low $x$ a transition from a dilute parton gas to 
a dense parton liquid takes place.
We derive geometrical scaling for the structure function in 
deep inelastic scattering at low $x$  from
a diverging correlation length
$\xi(x)$ of Wilson lines near the light cone.
QCD (SU(3)) in $2+1$  space - time
dimensions 
near the light cone becomes a critical theory in the limit of 
$x \rightarrow 0$ with a diverging correlation length
$\xi(x) \propto x^{-\frac{1}{2 \lambda_2}}$ where the exponent
$\lambda_2=2.52$ is obtained from the center group Z(3) of SU(3).
  
\end{abstract}

\maketitle

High energy electron proton scattering has presented exciting new
experimental results in QCD. The behaviour of cross sections 
with energy, a long standing issue in hadronic physics, has gained in
interest with the availability of small size probes. 
The color dipole in the photon can be made small by increasing the 
virtuality $Q$ of the photon.
In the course of $x$-evolution
the photon wave function develops many additional dipoles which in
general
 diffuse
into distance scales beyond the original size $1/Q$. This increase in
dipole density and/or  size of the photon wave function is generally
believed to be the origin of the increasing high energy cross section.
Perturbative QCD has been partially successful to 
explain low $x$ physics. In a recent paper \cite{1}
we have shown how DGLAP- evolution of a non perturbatively obtained
gluon distribution can lead to a successful description of the 
structure function data at HERA.

In this note we follow the very promising
Wilson line method outlined in ref.~\cite{4} in a Hamiltonian
framework near the light cone \cite{3}.
Our approach is all the way nonperturbative in contrast to the
standard perturbative approach in this field. We do not share the
assumption that even at high dipole density when the average transverse
distance between dipoles becomes small the hadronic state is
accessible to a simple perturbative treatment, since the overall
size scale is still large. For total cross sections the momentum
transfer is zero  and this overall size
matters. 

We consider it as an advantage that experiment helps us to
unravel
the badly understood dynamics of partons near the light cone. For
years there has been a considerable effort to model and investigate 
QCD in light cone coordinates. 
We have followed the approach with
near light cone coordinates which
smoothly interpolate between the Lorentz and light front
coordinates :
\BA
  x^t =x^{+} &=& \frac1{\sqrt2} 
       \left\{ \left(1 + \frac{\eta^2}{2} \right)
       x^{0} + \left(1 - \frac{\eta^2}{2} \right) x^{3} 
       \right\} ,\nonumber \\
          x^{-} &=& \frac1{\sqrt2} \left( x^{0}-x^{3} \right)~.
\label{Coor}
\EA

In this approach the question of quantum constraint equations does not
arise, since we treat the negative fermion energy states and
transverse
electric fields as independent degrees of freedom. The price to pay
for this treatment is high. The arbitrary constant $\eta$ 
which labels the nearness to the light cone appears in the 
Hamiltonian.  
For spectrum calculations it is cumbersome to have such a
parameter, since in QCD we have to extrapolate to the  continuum limit 
with the help of the renormalization group which  becomes difficult
in the presence of the extra parameter $\eta$.
For the
discussion of scattering the parameter $\eta$ presents 
an advantage since it allows to
folllow the evolution of the physics with increasing energy.
Consider high energy photon-proton scattering at small $x=Q^2/s$,
where $s=W^2$ is the cm energy squared. Using the photon vector
$q, q^2=-Q^2$ and the proton vector $p,p^2=m^2 \approx 0$ 
we can define two light-like vectors 
\BA
  e_1 &=& q- \frac{q^2}{2 pq}p\nonumber \\
      &=& q +x p\\
  e_2 &=& p.
\EA
For finite energies the vector of the photon $q$ can be calculated as
linear combination of the light-like vector $e_1$ with a small amount
of $e_2$ admixed

\BA
  e_{\eta} &=& q +xp -\frac{\eta^2}{2} p\nonumber \\
           &=& e_1-\frac{\eta^2}{2}e_2.
\EA 
One sees that in the limit of infinite energies the mixing $\eta$ 
is related to the 
Bjorken variable $x$ and vanishes as $\frac{\eta^2}{2}=x$.
Therefore it is natural to formulate high energy scattering
in near light cone coordinates. 
For small $x$ the eikonal phases
acquired by the quarks/antiquarks are the relevant collective variables.
The light cone Hamiltonian on the
finite light-like $x^-$ interval of length $L$ has Wilson line or
Polyakov operators
similarly to QCD formulated on a finite interval in imaginary time
at finite temperature
\begin{eqnarray}
{\cal U}({x}_\bot) &=& P \exp\left[ ig\int_0^L dx^-A_-(x_{\bot},x^-) 
\right] \\
P(\vec{x}_\bot) &=& \frac1{N_c} {\rm tr}~{\cal U}.
\label{Polyakov}
\EA

The dynamics of these
Polyakov operators is determined by the near light cone
Hamiltonian $H$ \cite {3}. 
In the $x \rightarrow 0$ limit, those pieces ${\cal H^{\eta} }$ 
of the Hamiltonian ${\cal H }$ dominate which
are most singular at $\eta=0 $ and do not 
couple to the three-dimensional
gauge fields $A_{\bot}$ and $\psi_+$.
These are the terms which are proportional to $\frac{1}{\eta^2}$
and of course also the parts where the conjugate momenta enter.
This reduced Hamiltonian contains all terms with the 
collective variables $a_-^{c_0}$ with the color indices $c_0=3,8$
\BE
a_-^{c_0}=\frac {1}{L} \int\limits_0^L    dx^-A_-^{c_0}(\vec{x}_\bot,x^-)
\EE
which determine the Wilson lines and live in a $2+1$ dimensional space.
We consider the dynamics of the fields
$a_-$ in vacuum, i.e. without the source term $e_{\bot}$. 
The $\eta-$ coordinates correspond to the physics in a
fast moving frame. Therefore, we factorize the reduced energy from the Lorentz boost
factor $\propto \frac{1}{\eta}$ and the  transverse lattice cut off $a$
\BA
\nonumber
h_{red} &=& 2 \eta a \int dx^- dx_{\bot} {\cal H^{\eta}}\\
        &=&\int  dx^- dx_{\bot} \sum_{c^0=3,8} 
\left( \frac{2 a}{\eta} tr(\frac{1}{L}e_{\bot}^{c_0}-
\nabla_{\bot}a_-^{c_0})^2+\frac{2 a \eta}{2 L^2}p_-^{c_0 \dagger}
p_-^{c_0}\right. \\
&-&\left. \frac{4 a}{\eta}\psi^{\dagger}_-~g a_-^{c_0}
\frac{\lambda^{c_0}}{2} \psi_-\right).
\EA
This Hamiltonian is accessible to a lattice treatment in a similar way
as the Hamiltonian in SU(2) ref. \cite{4}.
It can be rewritten in terms of 
zero mode fields $\varphi^{c_0}$:
\BA
\nonumber
\varphi^{c_0} (\vec b_\bot)&=&\frac{1}{2} g L a_-^{c_0}(b_{\bot}),\\
\frac{\delta}{\delta \varphi^{c_0}(b_{\bot})}&=&a^2 \frac{\delta}{\delta
\varphi^{c_0}(x_{\bot})}.
\EA

Large gauge transformations lead from one part of the fundamential
domain to another. The effective coupling constant of the Hamiltonian
coupling the zero mode fields $\varphi^{c_0}$ is 
$g^2_{\rm eff}=\frac{g^2 \eta L}{4 a}$.
In ref. \cite {4} we have done a Finite Size Scaling (FSS)
analysis for SU(2)
QCD obtaining
a se\-cond order transition as a function of the coupling $g^2_{\rm eff}$
between a phase with massive
excitations at strong coupling and a phase with mass less excitations at weak
coupling. In the strong coupling domain of $g^2_{\rm eff}$
the
energy of the rotators $\varphi^{c_0}$
is dominated by the electric energy 
$\propto g^2_{eff} \frac1J \frac\delta{\delta\varphi^{c_0}(\vec b)}
J \frac\delta{\delta\varphi^{c_0}(\vec b)}$ which corresponds to the
Laplacian in the group manifold. Each site
has an energy spectrum with a gap
$\eps_n=n(n+2)\eps_0$ in SU(2) or $\eps_n=n(n+4)\eps_0$
in SU(3). With decreasing  $g^2_{\rm eff}$ the magnetic coupling 
$\propto \frac{1}{g^2_{\rm eff}}(\varphi^{c_0}(\vec{b})
-\varphi^{c_0}(\vec{b}+\vec\eps))^2$
becomes stronger. A larger  nearest neighbor coupling 
leads to a coherently aligned  ground state which has
mass less excitations.
Consequently the mass gap vanishes at a sufficiently small $g^2_{eff}$.
The resulting critical SU(2) theory is in the
same universality class as the $Z(2)$ theory or the Ising model in 
$3$-dimensions, which has been checked  in the lattice simulations 
\cite {4} with the
available numerical accuracy.
In SU(3) the reduced Hamiltonian has rather  different symmetry 
properties
than the SU(2) Hamiltonian.  We think that  
the universality class of the reduced Hamiltonian in SU(3) is 
the three-state Potts model Z(3).
In each subregion of
the fundamental domain 
the zero mode variables $\varphi^3,\varphi^8$
are represented by one-spin orientation.
The relevant center group Z(3) has a  weak
first order transition whose critical line ends in a 
second order point in 
the presence of an external field. We conjecture that this external field 
is provided by the fermion zero mode density near the light cone.
To match the Hamiltonian lattice  with scattering the 
lattice constant $a$ is chosen to coincide with 
the photon resolution $ \approx \frac{1}{Q}$.
The longitudinal lattice extension $L$ must be larger than the color coherence length
of the $q \bar q $ state in the photon-proton
c.m. system. For details we refer to ref. \cite{2}. The
final conclusion is that
near the critical point the Wilson lines experience long range correlations
which means that dipoles 
in the photon wave function are correlated over
large distances. 
The correlation length $\xi$ increases
with $x \rightarrow 0$ as
\BE 
\xi \propto (\frac{x}{x_0})^{-\frac{1}{2 \lambda_2}} f_h(0).
\EE.

For finite correlation length there exists an 
intermediate range in transverse space ${1}/{Q}<x_\bot<\xi$ for which
the correlation function of Wilson lines is power behaved:
\BE
\label{e17}
\langle P(x_\bot) P(0)\rangle \approx 
\frac{1}{x_\bot^{1+n}}
\EE
where $n=0.04$ in Ising-like systems. This scaling region is responsible for the
well-known effect of critical opalescence in the gas liquid transition.
For larger distances $x_\bot>\xi$ the correlation
function decreases exponentially
\BE
\label{e18}
\langle P(x_\bot) P(0)\rangle \approx 
e ^{-{x_\bot}/{\xi}}.
\EE

We do not follow the small $x$ evolution  of the photon dipole
state, instead we give a qualitative description of the
effective photon size as a function of $x$ using the results of the
$2+1$ dimensional critical QCD SU(3) theory as a guiding principle.
We parameterize the dipole probability densities for the longitudinal 
and transverse photons,
\BA
\nonumber
\rho_{\gamma}^T&=&\frac{6 \alpha}{4 \pi^2}\sum_f  \hat e_f^2 \varepsilon^2
[z^2+
(1-z)^2]F_T(\eps x_\bot),\\
\rho_{\gamma}^L&=&\frac{6 \alpha}{4 \pi^2}\sum_f  \hat e_f^2 4 Q^2 z^2
(1-z)^2 F_L(\varepsilon x_\bot),\\
\varepsilon&=&\sqrt{Q^2 z(1-z)}.
\EA
The perturbative scale
for the dipole density is given by $\frac{1}{\varepsilon}$.
We modify  the photon wave function depending on the relation of the
correlation length $\xi$ of the Wilson loops to the perturbative scale
${1}/{\varepsilon}$.
We set 
\BE
\xi= \frac{1}{\varepsilon} (\frac {x}{x_0})^{-\frac{1}{2 \lambda_2}}\ .
\EE
For the 
reference Bjorken parameter $x_0=10^{-2}$
the correlation length is fixed at the perturbative scale. 
The critical exponent $\frac{1}{2 \lambda_2}=0.2$ determines
the Wilson line correlations for $x<x_0$.
If the transverse size of the dipole 
is smaller than the perturbative
length scale $x_{\bot}<\frac{1}{\varepsilon}$
we use  the  perturbative dipole densities 
$F_{T}(\varepsilon x_\bot)= K_{1}(\varepsilon x_\bot)^2 $ 
and
$F_L(\varepsilon x_\bot)= K_0(\varepsilon x_\bot)^2 $.
For ${1}/{\varepsilon}<x_\bot< \xi$ 
we modify  the perturbative dipole
densities using the correlation functions of the critical theory,
Eqs.~(\ref{e17},\ref{e18}),
\BA
\nonumber
F_{T/L}(\eps x_\bot) &=& K_{1/0}(\eps x_\bot)^2 ~~~~~~~~~~~~
\textrm{ for  } x_\bot <\frac{1}{\varepsilon}, \\
\nonumber
               &=& K_{1/0}(1)^2 (\frac{1}{\eps x_{\bot}})^{2+2 n} ~~~
\textrm{for  }  \frac{1}{\eps} < x_\bot < \xi ,\\
               &=& K_{1/0}(x_\bot/\xi)^2 (\frac{1}{\xi \eps})^{2+2 n}~~
\textrm{for  }  x_\bot > \xi.
\label{eftl}
\EA

The current discussion about geometrical scaling evolves around the
concept of saturation which has been carried over from the traditional 
unitarity behaviour of profile functions in impact parameter space.
Recall that the integral over the profile function gives the
total cross section.
When the profile function reaches the unitarity limit, the
target/projectile becomes
black, a further increase of the cross section can only be
reached
by an increase in transverse size of the profile function.
Now the concept of saturation is also applied to a dipole nucleon
cross section 
which becomes flat with increasing dipole size. The
Golec-Biernat cross section becomes flat at  
$r_{\bot}\approx1.0$ fm for $x=10^{-2}$. Since we consider all
the evolution to happen in the photon, this dipole nucleon
cross section is kept fixed as a function of energy, i.e as a
function of $x$.  In our picture
the increasing cross section comes from an increasing size of
the largest dipole state in the photon 
which is determined by the Wilson line correlation
function as given before. 
The fact that this effective density obeys approximate scaling
behaviour in the region $r_{\bot}<\xi$ is in our opinion the 
fundamental reason for geometric scaling. The photon-proton 
cross section can only depend on the ratio $\frac{R_0²}{\xi(x)²}$. 
We see this as follows: 
The effective dipole density combined with an energy independent
dipole-proton cross section determines the
structure function $F_2$ and the photon-proton
cross section.

\BA
F_2(x,Q^2) &=& \frac{Q^2}{4 \pi^2 \alpha }(\sigma_{\gamma p}^{T,tot}+
   \sigma_{\gamma p}^{L,tot}),\\
\sigma_{\gamma p}^{T/L,tot}&=&\int d^2 x_\bot \int _0 ^1 dz
\rho_{\gamma}^{T/L}(x_\bot,z) \sigma (x_\bot)
\EA
The Golec-Biernat-Wuesthoff \cite{5}
dipole-proton cross section is approximately 
equal to a simple quadratic function at small distances 
$r<2 R_0$ and a constant function for $r> 2 R_0$,
\BE
\sigma(r) \approx \sigma_0 \left (\frac {r^2}{4 R_0^2} \Theta(2 R_0- r) +
\Theta(r-2R_0) \right).
\EE

The numerical values $R_0=0.33$ fm at
$x_0=10^{-2}$ is independent of $x$.
One can demonstrate geometrical scaling of 
the photon-proton cross section rather simply.
We neglect 
the exponentially suppressed part of the dipole density in the
integral over large transverse distances and set 
the anomalous dimension $n \rightarrow 0$.
Then one gets  the dominant transverse
cross section in F$_2$ by integrating up to  the 
correlation length $\xi$, using the fact that 
$K_1(r \varepsilon)=1/(r \varepsilon)$ for
$r \varepsilon < 1$,
\BE
\sigma_T=\frac{3 \alpha}{\pi} \sum e_f^2 Q^2 \sigma_0\\
\int_0^1 dz(z^2+(1-z)^2)\int_0^{\infty} r dr \frac{z(1-z)}{r^2 \varepsilon^2}\\
\Theta(1-\frac{r^2}{\xi^2}) \frac{\sigma(r)}{\sigma_0}.
\EE

Redefining the integration variable as
$r'=r (x/x_0)^{\frac{1}{2 \lambda_2}} $ one obtains that the $\gamma^* -p$
cross section obeys geometrical scaling and depends only
on $R_0/\xi(x)=R(x) Q$ 
where
\BA
R(x)&=&R_0(\frac{x}{x_0})^{\frac{1}{2 \lambda_2}},\\
\frac{1}{2 \lambda_2} &=& 0.2.
\EA
The critical theory gives the phenomenologically obtained power dependence
of $R$ with $x$. 
The favored $x$-dependence of GBW is in the
range $0.145$ and $0.20$. The critical behavior  gives the power 
$\frac{1}{2 \lambda_2} = 0.2$. Without a model for the proton source, it is
not possible to obtain the absolute length $R_0$. 
The structure function scales as a function of
$Q^2 R_0^2 (x/x_0)^{{1}/{\lambda_2}}$ as can be easily
derived for the simplified dipole cross section. This geometrical
scaling
of the structure function is in good agreement with the data \cite{2}.

\vskip1cm
\noindent

\section*{ACKNOWLEDGEMENTS}

This work has been partially funded through the European TMR Contract
HPRN-CT-2000-00130 ``Electron Scattering off Confined Partons''
and supported by INTAS contract ``Nonperturbative QCD''. 

\end{document}